\documentclass[preprint2]{aastex}

\usepackage{epsf}  
\usepackage{amsmath}  
\shorttitle{Satellite Survival in CDM Cosmology} 
\shortauthors{Boily et al.}
\slugcomment{Submitted to ApJ, march 2004}

\begin{document} 
 
\newcommand{\getinptfile}
{\typein[\inptfile]{Input file name}
\input{\inptfile}}

\newcommand{\mysummary}[2]{\noi {\bf SUMMARY}#1 \\ \noi \sl #2 \\ \capline 
	\hspace{-.13in} \raisebox{.0in}{$\sqcap$} \rm }  
\newcommand{\mycaption}[2]{\caption[#1]{\footnotesize #2}} 
\newcommand{\capline}{\mbox{}\hrulefill}
\newcommand{\mysection}[2]{ 
\section{\uppercase{\normalsize{\bf #1}}} \def\junksec{{#2}} } %
\newcommand{\mychapter}[2]{ \chapter{#1} \def\junkchap{{#2}}  
\def\thesection{\arabic{chapter}.\arabic{section}}
\def\thesubsection{\thesection.\arabic{subsection}}
\def\thesubsubsection{\thesubsection.\arabic{subsubsection}}
\def\theequation{\arabic{chapter}.\arabic{equation}}
\def\thefigure{\arabic{chapter}.\arabic{figure}}
\def\thetable{\arabic{chapter}.\arabic{table}}
}
\newcommand{\mysubsection}[2]{ \subsection{#1} \def\junksubsec{{#2}} }
\def\thenote{\addtocounter{footnote}{1}$^{\scriptstyle{\arabic{footnote}}}$ }

\newcommand{\myfm}[1]{\mbox{$#1$}}
\def\spose#1{\hbox to 0pt{#1\hss}}	
\def\ltabout{\mathrel{\spose{\lower 3pt\hbox{$\mathchar"218$}} 
     \raise 2.0pt\hbox{$\mathchar"13C$}}}
\def\gtabout{\mathrel{\spose{\lower 3pt\hbox{$\mathchar"218$}}
     \raise 2.0pt\hbox{$\mathchar"13E$}}}
\newcommand{\ltsim}{\raisebox{-0.5ex}{$\;\stackrel{<}{\scriptstyle \backslash}\;$}}
\newcommand{\simlt}{\ltsim}
\newcommand{\simgt}{\gtsim}
%
\newcommand{\unit}[1]{\ifmmode \:\mbox{\rm #1}\else \mbox{#1}\fi}
\newcommand{\ze}{\ifmmode \mbox{z=0}\else \mbox{$z=0$ }\fi }

%
\newcommand{\boldv}[1]{\ifmmode \mbox{\boldmath $ #1$} \else 
 \mbox{\boldmath $#1$} \fi}
%
\renewcommand{\sb}[1]{_{\rm #1}}%
\newcommand{\expec}[1]{\myfm{\left\langle #1 \right\rangle}}
\newcommand{\mone}{\myfm{^{-1}}}
\newcommand{\half}{\myfm{\frac{1}{2}}}
\newcommand{\nth}[1]{\myfm{#1^{\small th}}}
\newcommand{\ten}[1]{\myfm{\times 10^{#1}}}
\newcommand{\abs}[1]{\mid\!\! #1 \!\!\mid}
\newcommand{\as}{a_{\ast}}
\newcommand{\asr}{(a_{\ast}^{2}-R_{\ast}^{2})}
\newcommand{\bvm}{\bv{m}}
\newcommand{\calf}{{\cal F}}
\newcommand{\calI}{{\cal I}}
\newcommand{\calm}{{v/c}}
\newcommand{\calminf}{{(v/c)_{\infty}}}
\newcommand{\calQ}{{\cal Q}}
\newcommand{\calR}{{\cal R}}
\newcommand{\calw}{{\it W}}
\newcommand{\co}{c_{o}}
\newcommand{\cs}{C_{\sigma}}
\newcommand{\cst}{\tilde{C}_{\sigma}}
\newcommand{\cv}{C_{v}}
\def\dbar{{\mathchar '26\mkern-9mud}}	
\newcommand{\deldelr}{\frac{\partial}{\partial r}}
\newcommand{\deldelR}{\frac{\partial}{\partial R}}
\newcommand{\deldeltheta}{\frac{\partial}{\partial \theta} }
\newcommand{\deldelphi}{\frac{\partial}{\partial \phi} }
\newcommand{\ddotrc}{\ddot{R}_{c}}
\newcommand{\ddotxc}{\ddot{x}_{c}}
\newcommand{\dotrc}{\dot{R}_{c}}
\newcommand{\dotxc}{\dot{x}_{c}}
\newcommand{\Estar}{E_{\ast}}
\newcommand{\grpsi}{\Psi_{\ast}^{\prime}}
\newcommand{\kboltz}{k_{\beta}}
\newcommand{\levi}[1]{\epsilon_{#1}}
\newcommand{\limaso}[1]{$#1 ( a_{\ast}\rightarrow 0)\ $}
\newcommand{\limasinfty}[1]{$#1 ( a_{\ast}\rightarrow \infty)\ $}
\newcommand{\limrinfty}[1]{$#1 ( R\rightarrow \infty,t)\ $}
\newcommand{\limro}[1]{$#1 ( R\rightarrow 0,t)\ $}
\newcommand{\limrso}[1]{$#1 (R_{\ast}\rightarrow 0)\ $}
\newcommand{\limxo}[1]{$#1 ( x\rightarrow 0,t)\ $}
\newcommand{\limxso}[1]{$#1 (\xs\rightarrow 0)\ $}
\newcommand{\ls}{l_{\ast}}
\newcommand{\Ls}{L_{\ast}}
\newcommand{\mean}[1]{<#1>}
\newcommand{\ms}{m_{\ast}}
\newcommand{\Ms}{M_{\ast}}
\def\nb{{\sl N}-body }
\def\nbt{{\sf NBODY2} }
\def\nb1{{\sf NBODY1} }
\newcommand{\nuoned}{\nu\sb{1d}}
\newcommand{\ra}{\rightarrow}
\newcommand{\Ra}{\Rightarrow}
\newcommand{\rc}{r_{c} } 
\newcommand{\Rc}{R_{c} } 
\newcommand{\res}[1]{{\rm O}(#1)}
\newcommand{\rnsa}{(r^{2}-a^{2})}
\newcommand{\Rnsa}{(R^{2}-a^{2})}
\newcommand{\rs}{r_{\ast}}
\newcommand{\Rs}{R_{\ast}}
\newcommand{\Rsa}{(R_{\ast}^{2}-a_{\ast}^{2})}
\newcommand{\sa}{\sigma } 
\newcommand{\sac}{\sigma_{c} } 
\newcommand{\sas}{\sigma_{\ast} } 
\newcommand{\sasp}{\sigma^{\prime}_{\ast}}
\newcommand{\saxs}{\sigma_{\ast} } 
\newcommand{\sech}{{\rm sech}}
\newcommand{\tff}{t\sb{ff}} 
\newcommand{\ti}{\tilde}
\newcommand{\trel}{t\sb{rel}}
\newcommand{\ts}{\tilde{\sigma} } 
\newcommand{\tss}{\tilde{\sigma}_{\ast} } 
\newcommand{\vcol}{v\sb{col}}
\newcommand{\vs}{v_{\ast}  } 
\newcommand{\vsp}{v^{\prime}_{\ast}}
\newcommand{\vxs}{v_{\ast}  } 
\newcommand{\xs}{x_{\ast}}
\newcommand{\xc}{x_{c} } 
\newcommand{\xistar}{\xi_{\ast}}
\newcommand{\rmd}{\ifmmode \:\mbox{{\rm d}}\else \mbox{ d}\fi }
\newcommand{\rmD}{\ifmmode \:\mbox{{\rm D}}\else \mbox{ D}\fi }
\newcommand{\valfven}{v_{{\rm Alfv\acute{e}n}}}

%
\newcommand{\noi}{\noindent}
\newcommand{\bc}{boundary condition }
\newcommand{\bcs}{boundary conditions }
\newcommand{\Bcs}{Boundary conditions }
\newcommand{\lhs}{left-hand side }
\newcommand{\rhs}{right-hand side }
\newcommand{\wrt}{with respect to }
\newcommand{\iras}{{\sl IRAS }}
\newcommand{\cobe}{{\sl COBE }}
\newcommand{\Oh}{\myfm{\Omega h}}
%
\newcommand{\etal}{{\em et al.\/ }}
\newcommand{\eg}{{\em e.g.\/ }}
\newcommand{\etc}{{\em etc.\/ }}
\newcommand{\ie}{{\em i.e.\/ }}
\newcommand{\viz}{{\em viz.\/ }}
\newcommand{\cf}{{\em cf.\/ }}
\newcommand{\via}{{\em via\/ }}
\newcommand{\apriori}{{\em a priori\/ }}
\newcommand{\adhoc}{{\em ad hoc\/ }}
\newcommand{\viceversa}{{\em vice versa\/ }}
\newcommand{\versus}{{\em versus\/ }}
\newcommand{\qed}{{\em q.e.d. \/}}
\newcommand{\<}{\thinspace}
%
\newcommand{\km}{\unit{km}}
\newcommand{\kms}{\unit{km~s\mone}}
\newcommand{\kmsa}{\unit{km~s\mone~arcmin}}
\newcommand{\kpc}{\unit{kpc}}
\newcommand{\mpc}{\unit{Mpc}}
\newcommand{\hkpc}{\myfm{h\mone}\kpc}
\newcommand{\hmpc}{\myfm{h\mone}\mpc}
\newcommand{\parsec}{\unit{pc}}
\newcommand{\cm}{\unit{cm}}
\newcommand{\yr}{\unit{yr}}
\newcommand{\au}{\unit{A.U.}}
\newcommand{\AU}{\au}
\newcommand{\gm}{\unit{g}}
\newcommand{\solar}{\myfm{_\odot}}
\newcommand{\solarm}{\unit{M\olar}}
\newcommand{\Lsun}{\unit{L\solar}}
\newcommand{\Rsun}{\unit{R\solar}}
\newcommand{\seconds}{\unit{s}}
\newcommand{\micro}{\myfm{\mu}}
\newcommand{\micrometer}{\micro\mbox{\rm m}}
\newcommand{\Mdot}{\myfm{\dot M}}
%
%
%
\newcommand{\dgr}{\myfm{^\circ} }
\newcommand{\ddgr}{\mbox{\dgr\hskip-0.3em .}}
\newcommand{\mnt}{\mbox{\myfm{'}\hskip-0.3em .}}
\newcommand{\scnd}{\mbox{\myfm{''}\hskip-0.3em .}}
\newcommand{\hr}{\myfm{^{\rm h}}}
\newcommand{\dhr}{\mbox{\hr\hskip-0.3em .}}
%
%
%
%
%
%
%
\newcommand{\refindent}{\par\noindent\hangindent=0.5in\hangafter=1}
\newcommand{\figpar}{\par\noindent\hangindent=0.7in\hangafter=1}
%
%

\newcommand{\mybiblio}{\vspace{1cm}
		       \setcounter{subsection}{0}
		       \addtocounter{section}{1}
		       \def\junksec{References} 
 }

%
%
%

%
%
%
%
%

\newcommand{\vol}[2]{ {\bf#1}, #2}
\newcommand{\jour}[4]{#1. {\it #2\/}, {\bf#3}, #4}
\newcommand{\physrevd}[3]{\jour{#1}{Phys Rev D}{#2}{#3}}
\newcommand{\physrevlett}[3]{\jour{#1}{Phys Rev Lett}{#2}{#3}}
\newcommand{\aaa}[3]{\jour{#1}{A\&A}{#2}{#3}}
\newcommand{\aaarev}[3]{\jour{#1}{A\&A Review}{#2}{#3}}
\newcommand{\aaas}[3]{\jour{#1}{A\&A Supp.}{#2}{#3}}
\newcommand{\jgeo}[3]{\jour{#1}{Journal of Geophysical Research}{#2}{#3}}
\newcommand{\rmp}[3]{\jour{#1}{Rev. Mod. Phys.}{#2}{#3}}
\newcommand{\science}[3]{\jour{#1}{Science}{#2}{#3}}
\newcommand{\vistas}[3]{\jour{#1}{Vistas in Astronomy}{#2}{#3}}

\newcommand{\leftb}{<\!\!} \newcommand{\rightb}{\!\!>} 
\newcommand{\new}[1]{ {\bf #1}}

\title{Satellite Survival in CDM Cosmology}  
\author{
C.~M. Boily\altaffilmark{1} }
\affil{Observatoire astronomique, 11 rue de l'Universit\'e, F-67000 Strasbourg,  France}
\email{cmb@astro.u-strasbg.fr}
\author{N. Nakasato\altaffilmark{2}}
\affil{Department of Astronomy, University of Tokyo, Bunkyo-ku, Tokyo 113-0033, Japan} 
\and
\author{ R. Spurzem \& T.~Tsuchiya\altaffilmark{3}}
\affil{Astronomisches Rechen-Institut, M\"onchhofstrasse 12-14,  D-69120 Heidelberg, Germany }

\altaffiltext{1}{Corresponding author}
\altaffiltext{2}{Current address:  Institute of Advanced Physical \& Chemical Research, 
2-1 Hirosawa, Wako-shi, Saitama 351-019, Japan}
\altaffiltext{3}{Now at SGI Japan, Ltd., Ebisu Garden Place Tower 31F,
  4-20-3 Ebisu, Shibuya-ku, Tokyo 150-6031, Japan}

\received{Received September 15, 1999; accepted March 16, 2000} 
 
   \begin{abstract}  
We study the survival of sub-structures (clumps) within larger  
self-gravitating  dark matter haloes. Building on scaling relations  
obtained from $N$-body calculations of violent relaxation, we argue 
that  the tidal field of galaxies and haloes can 
only destroy  sub-structures if spherical symmetry is imposed at formation.  
 We explore other mechanisms that may tailor the number of  
halo sub-structures during the course of virialisation. Unless the  
larger halo is built up from a few large clumps, we find that clump-clump  
encounters are unlikely to homogenise the halo on a dynamical timescale.  
 Phase-mixing would proceed faster in the inner parts and allow for the  
secular evolution of a stellar disc.    
   \end{abstract} 
  
%
 
\section{Introduction} 
  High-resolution simulations of structure formation in an  $\Lambda$CDM cosmogony have 
 revealed a large number of substructures within   
dark matter haloes (e.g. Ghigna et al. 2000; Moore 2001).  
  The mass function of these dark clumps (Moore et al. 1999a; de Lucia et al. 2004) 
  
\begin{equation} n(m) \propto m^{-2} \label{eq:clumpmf} \end{equation}  
in the mass range $10^8$ to $10^{11}$ solar masses.  These orbiting  
self-bound dark matter satellites would drag along baryonic 
 matter to a non-negligible fraction of their mass. Consequently  
galaxies should harbour a large number of dwarf galaxies, when only  
a handful are found (Kauffmann et al. 1993; Moore et al. 1999a; see Binney \& Merryfield 1999).  
  
The situation is made worse from a dynamical standpoint. Massive dark clumps would perturb the   
vertical structure of galactic discs through tidal heating, when their thin structure  
suggests the immediate neighbourhood  of discs is devoid of such perturbations  
(Toth \& Ostriker 1992; Moore et al. 1999a).  \new{The large-mass end of the 
clump  mass distribution function is robust in view of
Simulation  checks  find the large-mass end of the clump mass distribution function 
robust to numerical resolution issues (Moore et al. 1999b; Gao et al. 2004; Power et al. 2003).} Therefore, the orbital 
distribution of clumps in phase space  
must allow for \new{long periods of unperturbed evolution} by the disc, as  
emphasised by Navarro (2002) and Font et al. (2003).   A more severe problem is  
the narrowness of streams of stars associated with dwarf galaxies tidal debris  
(Ibata  et al. 2001; Johnston et al. 2001), which   
suggests a smooth background halo potential in order to preserve the cohesion of  
the stream, in direct  conflict with computer simulations of structure formation.  \new{Observational 
evidence drawn from solar neighbourhood kinematics points to a fine-meshing of low-mass streams 
to account for coherent motion in and out of the galactic plane (Helmi et al. 2002; Gould 2003).}
To settle this issue requires \new{both a precise map of the morphology of haloes and}  
 convergence in the mass distribution function of clumps with a resolution  
down to \new{the sub-dwarf mass range},   
still a challenge to present-day computer models (see e.g. Power et al. 2003). 
\new{The morphology of haloes has 
been discussed at length by several authors  (e.g., Moore 2001; Ghigna et al. 2000; Fukushige \& Makino 2003; Power et al. 2003; Navarro et al. 2004).  Several contributions in Natarajan (2002) give a broad overview of the formation processes and  equilibrium properties of haloes derived from numerical simulations and possible observational tests for their detection.}  \newline  
 
Particle-based calculations of galaxy formation proceed with a number routinely approaching  
a few $\times 10^7$ particles for a whole simulation. 
The mass of individual particles that make up dark haloes in these  
simulations $\simeq 10^6 M\solar$ is still large compared with a mass spectrum  
 of dark matter that may yet include a population  
of brown dwarfs (i.e., stellar masses).  \new{Furthermore,  of order $\sim 10^5$ mass elements must participate in the formation of a self-gravitating body to resolve the growth of potential energy adequately (Boily et al. 2002;  Roy \& Perez 2004), or a mass resolution of $10^8/10^5 \simeq 10^3\, {\rm M} \solar $ to account for dwarf-size structures.} This leaves a gap in mass resolution of some three orders of magnitude with that achieved by present-day simulations.  
 In this context, we need establish when the statistics of dark matter clumps derived from  
N-body computations may be scaled up to \new{actual} galactic systems. A crude picture  
of halo formation divides the process in two stages: one of rapid collapse 
on a free-fall time  scale, followed by a second, longer period of   
sporadic accretion (Bower 1991; Lacey \& Cole 1994; Zhao et al. 2003a,b). 
\new{The first stage involves structures spread over a narrow range of masses, while the 
second stage sees low-rate accretion by a dominating, central body.} 
The growth of potential energy during the free-fall   stage of formation sets the maximum  
phase-space density of the distribution function by efficiently, if incompletely,  
redistributing  binding energy between particles \new{(Lynden-Bell 1967; van Albada 1982)}.  
 
The growth of potential energy during collapse of self-gravitating systems  
is a function of the initial morphology  and the number of particles (or quivalently,  
mass resolution) used in the computation (e.g. Boily et al. 2002 \new{and references therein}).  
 A direct consequence of this  is that the maximum  
tidal field $\propto \nabla^2\phi(r)$  experienced by galactic satellites   \new{(dark or baryonic) 
as they fly through  the system}  varies  with  the morphology of the underlying  
halo, i.e., its formation history. For instance, at fixed resolution,  
 the strength of tidal fields reaches a  
 higher maximum for an initially spherical distribution 
 than for an axi-symmetric or triaxial one. The simplification of symmetric distributions  
allows to scale up the results with particle number exactly (see \S2 below).   
Would the tidal heating experienced by dark clumps during infall,  
scaled up to actual galactic halo particle numbers, be sufficient to un-bind them?  
In this short contribution, we apply the results \new{violent relaxation studies}   to the  
tidal heating of galactic substructures in a hierarchical Einstein-deSitter universe. We   
show that the tidal field developing during infall may yet be sufficient to erase  
sub-structures  smaller than a critical linear size $l_c$ if the particle number $N$ is sufficiently large  
{\it and} the galactic halo shows axial- or spherical symmetry during infall. 
 
\section{Scaling of tidal fields} 
Aarseth, Lin and Papaloizou (1988) (hereafter ALP+88) have  
shown that the growth rate of global modes of fragmentation during violent relaxation is such that  
the minimal radius achieved by a spherical distribution scales with the simulation particle number $N$
as  
 
\begin{equation} {\cal C} \equiv \frac{R(0)}{R(\tff)} \propto N^{1/3} \label{eq:C} \end{equation} 
where $R(0)$ is the initial system size enclosing $N$ identical mass elements and $R(\tff)$ the  
radial size at the free-fall time $\tff$ defined by  
 
\begin{equation} \tff = \sqrt{ \frac{3\pi}{32 G <\!\!\rho(a,0)\!\!> 
   } } \label{eq:tff} \end{equation}   
where $<\!\!\rho(a,0)\!\!> = 4\pi M/3 a^3 $ is the mean density inside the particle's initial radius, $a$.   
A recent study extended the result (\ref{eq:C}) empirically with N-body calculations   
to axisymmetric (cylindrical) and triaxial initial configurations  
(Boily et al. 2002). It was found in these cases that ${\cal C}$  scales with particle number  
as  
 
\begin{equation} {\cal C} \propto \begin{cases} N^{1/6} & \ {\rm (axisymmetry)} \\  
  \ {\rm constant} & \ ({\rm triaxial}). \end{cases} \label{eq:scaling}  
\end{equation}  
The factor ${\cal C}$ achieved by systems starting from triaxial distributions  
remains $\ltabout 40$ for $\simeq 86\%$  
 of the parameter space of axial ratios $a:b:c$, however the maximum achieved by individual 
realisations is highly  sensitive to the initial axes ratios and may yet diverge.   
N-body simulations using direct-integration algorithms  
(ALP+88; Theis \& Spurzem 1999; Boily, Clarke \& Murray 1999) or  
FFT integrators (e.g., Boily et al. 2002)  
  reproduce  these relations over several decades in particle number,  
giving confidence that the results  
are well recovered independently of the numerical method used. A consequence of (\ref{eq:C}) and (\ref{eq:scaling})  is that a spherical or axisymmetric  
 distribution will contract ever more as the  number  N of particles  is increased.  
 The morphological  
evolution of haloes and their substructures will not be well resolved if at formation the  
particle number which take part in the relaxation phase is too low.  
Zhao et al. (2003a,b) show that the formation history runs through a rapid accretion phase,  
followed by slow growth of the outer halo. The scaling law  
(\ref{eq:C}) and (\ref{eq:scaling})  obtained for  violent relaxation would therefore  
apply best to  the first, early phase of formation, but not the late one.    
 
\section{Cosmological application}   
The mean tidal field of a collapsing  halo is evaluated from the time-dependent second derivative of the  
gravitational potential  
 
\begin{equation} F_t = l_c \times\left.\nabla^2 \Phi\right|_R  = - l_c \,\frac{G M}{R^3}  
\label{eq:potential} \end{equation}  
with $r< $ the system radius $R$, and double-differencing  \wrt $r$ at 
fixed time yields a measure of the tidal field at $R$ acting on  a clump of size $l_c$.  
It is clear that the tidal force is unbound if during  collapse $R(\tff)\rightarrow 0$, as would occur e.g. for a large-N spherical distribution, cf. (\ref{eq:C}).  
 
Bound clumps would survive  violent relaxation  
 if their binding energy is higher than the tidal heating  
they suffer during that phase (van Albada 1982; Tsuchiya 1998).  
  As shown from (\ref{eq:potential}), remnant structures  
will  be severely disrupted  if the maximal potential depth  achieved is large.  
 To progress further we need invoke a result for structure formation in an 
 expanding Universe. Present-day data support an asymptotically flat  
metric for the Universe, and hence the  Einstein-deSitter  
cosmogony remains attractive.  
 
\subsection{Background tidal heating}  
In an EdS universe the relation between a bound structure's mass and 
virial radius is (Kaiser 1986; Padmanabhan 1993; Somerville \& Primack 1999)  
 
\begin{equation} R \propto M^\gamma \label{eq:RM} \end{equation}  
with $\gamma$ known in terms of the power-spectrum of density  
fluctuations $P(\boldv{k})$ at wavenumber $\boldv{k}$. 
The classic CDM power spectrum at the time $\tau$ of galaxy formation accounts for 
the time-evolution of structure from a bottom-up point of view. The relation between 
$P(\boldv{k}; \tau)$ and the Zeldovich spectrum arising from post-inflation decoupling, $P(\boldv{k}; t_i)$, 
is (see Bardeen et al. 1986, Appendix G): 
\begin{equation} P(\boldv{k}; \tau) = \left[ \frac{a(t_i)}{a(t)}\right]^2\, T^2(\boldv{k}) P(\boldv{k}; t_i) 
\end{equation} 
where $a$ is the cosmical expansion factor and $T$ the tranfer function which is well approximated 
 analytically by 

\begin{equation} T(q) = \frac{\ln ( 1+2.34 q) }{2.34q} \, \left( 1 + 3.9 q + 259q^2 + 163q^3 + 2027 q^4\right)^{-\displaystyle{\frac{1}{4}}} \label{eq:transfer}\end{equation} 
with $q \equiv |k|\sqrt{\theta}/( \Omega h^2 {\rm Mpc}^{-1})$, $\theta = \rho_{rel}/1.68 \rho_\gamma$ is the ratio of 
relativistic paticles to photon energy densities. On the largest scales the Zeldovich spectrum $P(\boldv{k}; \tau) \propto P(\boldv{k}; t_i) \propto k $  is recovered from (\ref{eq:transfer}), while on small scales 
$P(\boldv{k}; \tau) \propto k^{-3}$.
If we fit the power spectrum locally to a power-law of index $n_k$, 
$P(\boldv{k}) \propto k^{n_k}$, the index shifts progressively 
 from $n_k = +1$  to $n_k = -3$ as we explore smaller scales. 
The indices $n_k$ and $\gamma$ are linked through the mass-radius relation (\ref{eq:RM}) 
\[\gamma = (n_k +5)/6\, .\] 
Note that $\gamma$ is related to the power-index $\nu$ relating mean  
density and mass, $\rho \propto M^\nu$, by $\nu = \gamma - 3 = (n_k - 13)/6 $;  
the free-fall time $\tff \propto 1/\sqrt{\rho}$ therefore scales with mass as   
\[ \tff \propto M^{(13-n_k)/12}.\] 
On the smallest scales $n_k \rightarrow -3 $ and $\tff \propto M^{4/3}$, a steeper  
relation than on the largest scales when $n_k \rightarrow +1$ and $\tff \propto M$.  
Therefore small clumps have fully virialised when the violent relaxation  
phase of the larger halo begins.  
 
 Since there is no fixed scales of mass or radius in gravitational dynamics, all 
virialised structures obey the same relation (\ref{eq:RM}). If we lump  
together all those of virial radius $< r$, of mass $m$ chosen such 
that $M/m = N \gg 1$, we have  
 
\begin{equation} \frac{R}{r} \propto \left( \frac{M}{m}\right)^{\gamma} = 
N^{(n_k+5)/6} \ .\label{eq:Rr} \end{equation}  
In a hierarchical universe, small structures form first and hence the 
$N$ small clumps have virialised well before the large underlying 
halo, of total mass $M$.  At constant mass the virial 
theorem  provides a relation between equilibrium radius an initial size $R_i$  
for a self-gravitating system:  
 
\begin{equation} R_i = 2R \label{eq:rvir} \end{equation}  
which applies equally to all structures. We need relate the tidal 
field (\ref{eq:potential}) with (\ref{eq:C}) and (\ref{eq:scaling})  
 to determine whether a  
structure of size $r_i$ survives the formation of the larger halo, of 
radius $R_i$. The energy transferred to a small clump by tidal forces  
during infall is adequately quantified by Spitzer's (1958) impulse 
approximation formula even for relatively slow encounters (Aguilar \& 
White 1985). This gives confidence that it will hold in the present 
context, where velocities are high. Tsuchiya (1998) finds this to be  
correct in his study of relaxing Plummer distributions. The tidal 
energy gained $\Delta E$ by a substructure of mass $m$, radius $r$, 
and internal binding energy $E$ may be evaluated for a single passage 
at velocity $V$  
across the background halo (or galaxy) potential at the time of 
collapse, when the tide is maximum. In the impulse approximation, this 
is given by  
 
\begin{equation} \Delta E = \frac{4G^2M^2m}{3R(\tff)^4 V^2}\, 
\bar{r}^2  \approx \left(\frac{\bar{r}}{R_i}\right)^3 {\cal C}^3 \, 
\frac{M}{m}\ E\,. \label{eq:dEtide} \end{equation}  
Note that (\ref{eq:dEtide}) applies to the tidal field of the  
background potential, and does not account for individual clump 
encounters. These are discussed later. Substituting (\ref{eq:Rr}) in  
(\ref{eq:dEtide}) we get  
 
\begin{equation} \Delta E = N^{-(n_k+3)/2} \left( \frac{{\cal 
C}}{2}\right)^3 \, E \label{eq:dEn} \end{equation}  
whence we deduce that $\Delta E \ll E $ if the coefficient 
$N^{-(n_k+3)/2} {\cal C}^3$ remains small. For spherical systems the  
collapse factor ${\cal C}$ obeys (\ref{eq:C}) and therefore  
 
\[ \frac{\Delta E}{E} \propto N^{-(n_k+1)/2} \hspace{1cm} {\rm (spherical\  
distributions)} \] 
while for axisymmetric or triaxial distributions we find from (\ref{eq:scaling})  
 
\begin{equation} \frac{\Delta E}{E} \propto \begin{cases}  
N^{-(n_k+2)/2} & 
{({\rm spheroidal})} \\ 
 N^{-(n_k+3)/2} & {{\rm (triaxial)}}.\end{cases}  
 \end{equation}  
The consequences of these results in  
relation to the power-spectrum of density fluctuations is 
clear: for clumps orbiting in a collapsing spherical halo or galaxy, 
tidal heating will be ineffective provided $n_k > -1$. If the  
underlying distribution is axi-symmetric, tidal heating will be  
ineffective when  $n_k > -2$.  However for  triaxial  
initial conditions this will hold true if $n_k > -3$.  
 Since the  power-spectrum of observed matter distribution is never  
steeper than $n_k = -3$, we deduce that the bulk of substructures or  
clumps evolving in larger structures, such as dark matter haloes,  
 will survive the formation of 
triaxial larger structures, if gravity alone fixes their binding energy.  
 
\subsection{Tidal heating due to other substructures}  
The above conclusion only concerns the response of clumps to the 
background tidal field. We may consider the interaction between clumps  
themselves as they cross the dense system. To this end we consider 
the tidal heating by two equal-mass substructures during an encounter. Substituting  
$M\rightarrow m$ in (\ref{eq:dEtide}) and the radius $R(\tff)$ by the  
mean distance  between clumps $x \approx R(\tff)/N^{1/3}$, an effective impact  
parameter, we find  
 
\[ \Delta E = \frac{4G^2m^3}{3(R/N^{1/3})^4 V^2}\, 
\bar{r}^2  \approx \frac{2}{3} \frac{m}{M} N^{4/3} 
\left(\frac{\bar{r}}{R}\right)^3 \ E\,.  \] 
 
In the above we used $V^2 \approx 2GM/R$, with $M = N m$ the total 
system mass as before.  The expression reduces to  
 
\begin{equation}  
\Delta E \approx \frac{2}{3} N^{1/3} \left(\frac{\bar{r}}{R}\right)^3 \ E\,.   
\label{eq:clumps}  
\end{equation}  
Clearly to achieve $\Delta E \approx E$ requires $r \sim R$ or $N\gg 1$.  
 we may simplify (\ref{eq:clumps}) by  
substituting for $\bar{r}$ using (\ref{eq:Rr}). We then find  
 
\begin{equation}  
\Delta E <  \frac{2}{3} N^{-(13+ 3n_k)/6} \ E\,.   
\end{equation}  
Thus for any appreciable number $N$ the tidal heating due to encounters  
between clumps will be significant if $n_k < - 13/3 = - 4 \frac{1}{3}$.  
No regime of the power spectrum covers that range and hence encounters  
between clumps never produce significant tidal heating.   \new{A direct consequence of this 
is that while haloes form at different redshifts and sample  different regimes of the 
structure power-spectrum,  the mass distribution function of substructures should be 
robust against  cut-offs or significant changes to its shape. It is not clear yet whether the 
scale-free nature of the clump mass function (\ref{eq:clumpmf}) measured in N-body simulations 
(e.g. Ghigna et al 2000) can be extended to very small masses (see Gao et al. 2004).} 
 
There is of course one situation when substructures can heat-up  one  
another through tidal forces, which is when $\bar{r} \sim R(0) $ or $R(\tff)$  and $N \sim 1$.  
Indeed when $\bar{r}$ matches the mean inter-clump distance $\approx R(\tff)/N^{1/3}$,  we compute $\Delta E/E \sim 1$, always. However this situation is more appropriate to galactic mergers than the  
formation of halo through accretion of several sub-units, as would occur in any bottom-up calculation 
of galaxy formation. 

The survival of substructures is in part due to the large  
relative velocity $V$ established under the mutual potential of all 
clumps. Survival during in-fall is no guarantee that the substructures would  
remain once the halo has achieved virial equilibrium.  
To estimate more precisely the net rate  
of heating on one clump due to the background tidal field  as it 
crosses the system is made difficult 
because of  the large changes in potential taking place during violent relaxation. 
This is best done with numerical N-body calculations tailored for this 
problem. The very high-resolution simulations performed by e.g. Ghigna et 
al. (2000)  demonstrate the likely survival of most substructures in and 
around dark matter haloes post-virialisation, in support of the basic argument outlined here.  
 
\section{Discussion and conclusion} 
Galactic satellites survive the formation phase of triaxial haloes and galaxies and will  
be destroyed on long timescales as they orbit the host galaxy.  
 This result is obtained both from large N-body simulations (see also e.g. Bullock et 
al. 2000; Ghigna et al. 2000) and from \new{semi-analytic arguments (Moore et al. 1996; Taffoni et al. 2003)} as well as the fluid calculation presented here, and 
is therefore robust. 
 The initial conditions and subsequent evolution of N-body computer simulations still  
plague their interpretation and  application to observed galaxies in terms 
of simple estimates of satellite disruption times. For instance,  
Font et al. (2003) and Ardi et al. (2002) have questioned the rate of disc heating by in-falling dark satellites.     These authors find that thin discs may yet remain stable despite a high count of 
bound dark matter clumps, provided the clumps  do not follow near-radial orbits. The problem of 
disc heating would seemingly not occur if the inner region of the halo were isotropic in phase-space. 
We already noted that the destruction of dark satellites would be more effective in the deep potential of the 
inner halo. \new{Recently Gao et al. (2004) have re-derived statistics of halo substructures in computer simulations and found them to be less concentrated than the host halo. This and  
the coherence  of cold-stream satellite debris cold-stream debris could be interpreted in 
the light of the present analysis as pointing to a
near-spherical (and hence destructive), early phase of halo formation. Several studies 
have argued 
for a more spherical morphology in the inner region of haloes (e.g., Blumenthal et al. 1986; Dubinsky 1994). The 
cooling of baryons at the heart of DM haloes would provide a mechanism for this by locking the inner halo morphology to a  rounder shape than obtained from strictly gravitational evolution (Dubinsky 1994; Frenk et al. 1996). The inner morphology of galactic haloes would not 
automatically be spherical  if baryons have had time to cool and form discs {\em before} the 
halo assembly is comlete : when that is the case, the halo's inner morphology in equilibrium is even
 more sensitive to the formation history and depends for instance on the orientation of the 
 discs as they merge (see  Kazantzidis et al. 2004)} \newline 
 
Possible mechanisms that may disrupt dark clumps on the dwarf galaxy mass-scale and below  
include  supernova blow outs through gas irradiation and expulsion  
(Efstathiou 1992; Somerville 2002; Gnedin \& Zhao 2002).  
The net effect of gas loss on gravitationally bound  structures is unlikely  
to be effective  if the gas mass fraction  
is small (Hills 1980; Boily \& Kroupa 2002). Gnedin \& Zhao (2002) have argued that the peaked density  
profiles obtained from CDM numerical calculations would resist rapid removal of the gas  
under any realistic circumstances. Thus unless the \new{dwarf-size} 
clumps contain a very large fraction of baryonic matter they will survive any degree of gas heating. 
 C\^ot\'e et al. (2002) presented  a Monte Carlo  
 simulation of chemical enrichment of galaxies through gas-evaporation from clumps of dark matter initially seeded with baryons (uniform M/L ratio). 
They find that the mass function of seeded clumps required to match  
their sample of galaxies (in terms of chemical gradients and observable dwarf galaxy and star cluster  
populations) is similar to the mass function (\ref{eq:clumpmf}) obtained from large-N cosmological  
 calculations. This would suggest that the high-mass end of the clump mass function survives the  
formation of the host dark halo to produce the observed population of dwarfs. It does not however  
suppress the number of dark clumps that may still be orbiting the halo. 
Another route to solving the over-abundance of dwarf galaxies is by \new{preventing  (bright) baryons from forming stars}.  This can be achieved either through background UV radiation (Efstathiou 1992; Somerville 2002), or by preventing a Toomre instability from developing fully (Verde et al. 2002), 
effectively shutting off the formation of stars in the first place. \new{Gas-rich dwarfs would undergo 
substantial morphological evolution through ram stripping from the IGM ; their long-term fate (destruction or survival) must 
account for such evolution since it will change the dwarf binding energy through dissipation (Mayer et al. 2001).} 
\newline 
 
Dynamical friction can in principle provide an alternative  solution 
if the clumps spiral in rapidly and lose mass owing to tidal heating  
(Syer \& White 1998; Tormen et al. 1998).  
Computer simulations and analysis of decaying satellites show that a heavy satellite loses up to  
90\% of its mass in a few orbital periods (e.g. Klessen \& Kroupa 1998; Pe$\tilde{\rm n}$arrubia et al. 2002; Taffoni et al. 2003; see also Hashimoto et al. 2003).  
 van Kampen (2002) has argued that the effect of dynamical friction may yet be underestimated  
in computer simulations due to limited mass resolution.  Dynamical  friction, in conjunction with  
tidal forces, will cause the disruption of a satellite after 
a period of time  (e.g. Ibata et al. 1994; Klessen \& Kroupa 1998; Bullock et 
al. 2001). Bullock et al. (2001) argue that the halo stellar population of the Milky Way may be  
accounted for if sufficient satellite dwarf galaxies had already accreted their 
mass at the time the galactic halo formed and are then stripped of their less-bound stars by galactic tides.   
 Clearly the link between halo morphology, halo sub-structure statistics and stellar populations 
 offers more avenues for future work. \newline 
 
\acknowledgements  We acknowledge discussions with Rachel Somerville. 
This work  was funded in part by the   
      {Sonder For\-schungsbereich (SFB)\/} 439 programme in Heidelberg. 
An E.G.I.D.E.  travel grant awarded to CMB by the French Minist\`ere 
des Affaires \'Etrang\`eres is gratefully acknowledged.  

 

%
%


\end{document}